\documentclass[conference]{IEEEtran}
\IEEEoverridecommandlockouts
\usepackage{amsfonts}
\usepackage{amsmath}
\usepackage{balance}
\usepackage{cite}
\usepackage{hyperref}

\usepackage{ifpdf}
\ifCLASSINFOpdf
   \usepackage[pdftex]{graphicx}
   \graphicspath{{./pdf/}}
   \DeclareGraphicsExtensions{{.pdf}}
\else
   \usepackage[dvips]{graphicx}
   \graphicspath{{./eps/}}
   \DeclareGraphicsExtensions{{.eps}}
\fi

\usepackage[usenames]{color}

\begin{document}
\hyphenation{multi-symbol}
\title{ Analysis of Multi-Cell Downlink Cooperation \\ with a Constrained Spatial Model}
\author{\IEEEauthorblockN{ Salvatore Talarico,\IEEEauthorrefmark{1} Matthew C. Valenti,\IEEEauthorrefmark{1} and Don Torrieri\IEEEauthorrefmark{2}}
\IEEEauthorblockA{\IEEEauthorrefmark{1}West Virginia University, Morgantown, WV, USA. \\
\IEEEauthorrefmark{2}U.S. Army Research Laboratory, Adelphi, MD, USA.}
}
\date{}
\maketitle

\vspace{-1cm}
\thispagestyle{empty}

\begin{abstract}
Multi-cell cooperation (MCC) mitigates intercell interference and improves throughput at the cell edge.  This paper considers a cooperative downlink, whereby cell-edge mobiles are served by multiple cooperative base stations.  The cooperating base stations transmit identical signals over paths with non-identical path losses, and the receiving mobile performs diversity combining. The analysis in this paper is driven by a new expression for the conditional outage probability when signals arriving over different paths are combined in the presence of noise and interference, where the conditioning is with respect to the network topology and shadowing. The channel model accounts for path loss, shadowing, and Nakagami fading, and the Nakagami fading parameters do not need to be identical for all paths. To study performance over a wide class of network topologies, a random spatial model is adopted, and performance is found by statistically characterizing the rates provided on the downlinks.   To model realistic networks, the model requires a minimum separation among base stations.   Having adopted a realistic model and an accurate analysis, the paper proceeds to determine performance under several resource-allocation policies and provides insight regarding how the cell edge should be defined.

\end{abstract}

\section{Introduction} \label{Section:Intro}
One of the main factors limiting the throughput in cellular networks is intercell interference. A promising approach for mitigating such interference is {\em multi-cell cooperation}  (MCC), which involves the joint processing of signals transmitted to multiple base stations and/or received by them \cite{gesbert:2010,simeone:2011,marsch2011}. In practice, multi-cell cooperation is made possible by the presence of an infrastructure of backhaul links connecting individual base stations to a central processor or to one another \cite{irmer:2011}.  The LTE-Advanced cellular standard can implement such backhaul links through the X2 interface \cite{ltea}.

This paper presents a new and precise analysis for the MCC downlink.  The analysis is driven by a new closed-form expression for the conditional outage probability when signals arriving over different paths are combined in the presence of noise and interference, where the conditioning is with respect to the network topology and shadowing. The expression is a generalization of one recently presented by the authors in \cite{torrieri:2012} for the case of a single source transmitter, several interferers, and noise.  As in \cite{torrieri:2012}, the channel model accounts for path loss, shadowing, and Nakagami fading, and the Nakagami fading parameters do not need to be identical for all links.  The generalization is to allow for multiple source transmissions, which are diversity combined at the receiver.  Such processing is typical of a rake receiver that performs maximal-ratio combining of the resolvable multipath components of a direct-sequence code-division multiple access (DS-CDMA) downlink signals.  While this paper focuses on a DS-CDMA downlink for illustrative purposes, the results can be generalized to an orthogonal frequency-division multiple access (OFDMA) downlink with resolvable multipath components.

Whereas the conditional-outage-probability equation derived in this paper characterizes the performance of a particular topology, it is desirable to gain insight into performance for an entire class of topologies.  To study performance over a wide class of network topologies, a random spatial model is adopted, and performance is found by statistically characterizing the rates provided on the downlinks.  Particular metrics of interest include the {\em area spectral efficiency}, which is the average number of successful receptions per unit area, and the cumulative distribution of the rates provided on the downlinks, which characterizes the fairness of the system by identifying the fraction of users that are able to achieve a specified rate.  To model realistic networks, the model imposes a minimum separation among base stations.

Other authors apply spatial models to the conventional (non-cooperative) cellular downlink \cite{andrews:2011} and MCC cellular downlink \cite{xia:2012,huang:2013} and model the base-station locations as a realization of a random point process, thereby allowing the use of analytical tools from stochastic geometry \cite{stoyan:1996}.  The approach allows for the analytical characterization of performance metrics, such as coverage probability of a typical mobile user, by combining the effects of random base-station location, fading, and shadowing into a single random variable.  In particular, the spatial model in \cite{andrews:2011,xia:2012,huang:2013} assumes that the base stations are drawn from a two-dimensional Poisson point process (PPP) and that the network extends infinitely on the Euclidian plane.  In contrast, our model differs from those in the related literature because it permits a minimum spacing between base stations and limits the number of base stations in a finite geographic region.  This model is more realistic that those based on a pure PPP, which can result in base stations that are unrealistically closely spaced and an unbounded number of base stations within a given area.

A similar constrained spatial model was applied to a conventional non-cooperative DS-CDMA downlink in \cite{Valenti:2013}, and the main goal of the present paper is to extend the analysis and model of \cite{Valenti:2013} to the MCC downlink.   As in \cite{Valenti:2013}, the spatial model places a fixed number of base stations within a region of finite extent and enforces a minimum separation among the base stations.  The model for base-station placement is a binomial point process (BPP) with repulsion,   which we call a {\em uniform clustering} model \cite{torrieri:2012}.

Having adopted a realistic model and accurate analysis, this paper proceeds to determine performance under several resource allocation policies, and provides insight regarding how the cell edge should be defined.  Policies for rate control, power control, and cooperative cell association are considered and optimized with respect to the area spectral efficiency.

The remainder of this paper is organized as follows. Section \ref{Section:SystemModel} presents the signal model, culminating in an expression for the signal-to-interference-and-noise ratio (SINR).  Section \ref{Section:Outage} derives a closed-form expression for the outage probability, conditioned on the network topology.  Section \ref{Section:Policies} discusses policies for power allocation, rate control, and cell association.  Section \ref{Section:SpatialModel} presents a constrained random spatial model, and discusses metrics for statistically characterizing the performance of a class of topologies.  Section \ref{Section:Results} gives some example numerical results, comparing the alternative resource allocation policies.  Finally, the paper concludes in Section \ref{Section:Conclusion}.

\section{Signal-to-Interference-and-Noise Ratio} \label{Section:SystemModel}
The network comprises $M$ cellular base stations $\{X_1, ..., X_M\}$ and $K$ mobiles $\{ Y_1, ..., Y_K\}$ placed in a region of area $A_\mathsf{net}$.  The variable $X_i$ represents both the $i^{th}$ base station and its location, and similarly, $Y_j$ represents the $j^{th}$ mobile and its location.  Each mobile is served by one or more base stations. Let $\mathcal X_j$ be the set of base stations that serve mobile $Y_j$, and let $N_j = | \mathcal X_j |$ denote the number of base stations that serve mobile $Y_j$. Let $\mathcal Y_i$ be the set of mobiles connected to base station $X_i$, and $K_i = | \mathcal Y_i |$ be the number of mobiles served by $X_i$.  Furthermore, let $\mathcal G_j$ be the set of the indexes of the base stations serving $Y_j$ so that $Y_j \in \mathcal Y_i$ if $i \in  \mathcal G_j$.  If $Y_j$ cannot connect to any base station, which is possible when all nearby cells run out of available channels, then $\mathcal G_j = \emptyset$.

The downlink signals use orthogonal DS-CDMA sequences with a common spreading factor $G$. A rake receiver is used at each mobile, and the signals from the serving base stations are assumed to be separated in time by at least a chip period, denoted by $T_c$. The receiver at mobile $Y_j$ coherently combines all the resolvable multipath components from the set of serving base stations $\mathcal X_j$.
Because the sequences transmitted by a particular base station are orthogonal, the only source of intracell interference is due to the unresolvable multipath components and the corresponding loss of orthogonality. However, if the ratio of the maximum-power and minimum-power unresolvable multipath components is sufficiently small, then the unresolvable multipath components will have negligible effect.  For this reason, we neglect the intracell interference and assume that intercell interference is the only source of interference.

The signal is transmitted by base station $X_i$ to mobile $Y_{j} \in \mathcal Y_i$ with average power $P_{i,j}$.  We assume that the base stations transmit with a common power $P_0$ such that
\vspace{-0.1 cm}
\begin{eqnarray}
\frac{1}{1-f_p}
\sum_{j: Y_j \in \mathcal Y_i}
P_{i,j}
& = &
P_0
\label{eqn:pwr_constraint}
\end{eqnarray}
for each $i$, where $f_{p}$ is the fraction of the base-station power reserved for pilot signals needed for synchronization and channel estimation.

Using spreading sequences with a spreading factor of $G$ directly reduces the power of the intercell interference. While the intracell sequences transmitted by the serving base stations are synchronous, the varying propagation delays from the other base stations cause the intercell interference to be asynchronous.  Because of this asynchronism, the intercell interference is further reduced by the chip factor $h(\tau_{i,k})$, which is a function of the chip waveform and the timing offset $\tau_{i,k}$ at the mobile $Y_j$ between the signal received from interfering base station $X_k, \forall k \notin \mathcal G_j$, and the signal received from serving base station $X_i, \forall i \in \mathcal G_j$ \cite{torrieri:2011}.  It is assumed that $G/h(\tau_{i,k})$ is a
constant equal to $G/h$, and a common value of $h=2/3$ is adopted.  It follows that the despread power of an intercell interferer is attenuated by a factor of $G/h$, which we call the {\em effective} spreading factor, while despreading does not significantly affect the power of the desired signal.



Let $\rho_{i,j}$ represent the instantaneous received power of $X_i$ at mobile $Y_j$ after despreading, which
depends on the path loss, shadowing, fading, and effective spreading factor.
We assume that the path loss has a power-law dependence on distance.   In particular, for a distance $d \geq d_{0}$, the path-loss function is expressed as the attenuation power law
\vspace{-0.1cm}
\begin{equation}
f\left(  d\right)  =\left(  \frac{d}{d_{0}}\right)  ^{-\alpha}
\label{eqn:pathloss}%
\end{equation}
where $\alpha\geq2$ is the attenuation power-law exponent, and it is assumed
that $d_{0}$ is sufficiently large that the signals are in the far field.

Let $d_{i,j} = ||X_i-Y_j||$ be the distance between base station $X_i$ and mobile $Y_j$.  The instantaneous despread power of the signal from a serving base station ($i \in \mathcal G_j$) is
\begin{eqnarray}
   {\rho}_{i,j}
   & = &
   P_{i,j} g_{i,j} 10^{\xi_{i,j}/10} f( d_{i,j} ) \label{despread_1}
\end{eqnarray}
where  $g_{i,j}$ is the power gain due to fading and $\xi_{i,j}$ is a \textit{shadowing factor}.  The
$\{ g_{i,j} \}$ are independent with unit-mean, and $g_{i,j}=a_{i,j}^{2}$,
where $a_{i,j}$ is Nakagami with parameter $m_{i,j}$.
While the $\{g_{i,j}\}$ are independent from mobile to mobile, they are not
necessarily identically distributed, and for each mobile each link between $Y_j$ and $X_i$ can be characterized by a distinct Nakagami parameter $m_{i,j}$.
When the channel between $X_i$ and $Y_j$ experiences Rayleigh fading, $m_{i,j}=1$
and $g_{i,j}$ is exponentially distributed. It is assumed that the \{$g_{i,j}\}$ remain fixed for the duration of a time interval, but vary independently from interval to interval.  In the presence of log-normal shadowing, the $\{\xi_{i,j}\}$ are i.i.d. zero-mean Gaussian with variance $\sigma_{s}^{2}$. In the absence of shadowing, $\xi_{i,j}=0$.

The signal received from an interfering base station ($i \notin \mathcal G_j$) is further attenuated by the effective processing gain $G/h$, hence, its instantaneous despread power is
\vspace{-0.1 cm}
\begin{eqnarray}
   {\rho}_{i,j}
   & = &
   \frac{h}{G} P_{i,j} g_{i,j} 10^{\xi_{i,j}/10} f( d_{i,j} )  \label{despread_2}
\end{eqnarray}

The shadowing can be considered to be a random displacement between the base station and receiver.  Define the {\em effective distance} as
\vspace{-0.2 cm}
\begin{eqnarray}
\tilde{d}_{i,j}
&  = &
 10^{-\xi_{i,j}/(10\alpha)} d_{i,j}  \label{eff_dist}
\end{eqnarray}
which is the distance perturbed by the shadowing.  In the absence of shadowing, $d_{i,j} =\tilde{d}_{i,j}$.  Substituting (\ref{eff_dist}) into (\ref{despread_1}) and (\ref{despread_2}) gives the instantaneous despread power
\vspace{-0.1cm}
\begin{eqnarray}
\hspace{-0.5cm} \rho_{i,j} \hspace{-0.3cm}
& = & \hspace{-0.3cm}
\begin{cases}
{P}_{i,j} g_{i,j} f( \tilde{d}_{i,j} )  & \mbox{if $  i \in \mathcal G_j$} \vspace{0.2cm} \\
\left( \frac{h}{G} \right) {P}_{i,j} g_{i,j} f( \tilde{d}_{i,j} ) & \mbox{if $i \notin \mathcal G_j$}
\end{cases}
\label{eqn:power}%
\end{eqnarray}

If the receiver of mobile $Y_j$ is able to resolve the signal received from each base station whose indices are in the set $\mathcal G_j$, estimate the corresponding complex-valued channel gains, and reject the path crosstalk, then it may perform {\em maximal-ratio combining} (MRC) of the paths \cite{torrieri:2011}.  The resulting instantaneous SINR at mobile $Y_j$ by using (\ref{eqn:power}) and (\ref{eqn:pathloss}) is
\vspace{-0.1cm}
\begin{eqnarray}
\gamma_j
&  = &
\frac{\displaystyle{\sum_{i \in \mathcal G_j} g_{i,j}\Omega_{i,j}}}
{\displaystyle\Gamma^{-1}
+
\frac{h}{G}
\sum_{ i \notin \mathcal G_j }
g_{i,j}\Omega_{i,j}}
\label{Equation:SINR2}
\end{eqnarray}
where $\Gamma=d_{0}^{\alpha} N_j P_0 /\mathcal{N}$ is the signal-to-noise ratio
(SNR) at a mobile located at unit distance when fading and
shadowing are absent, $\mathcal{N}$ is the noise power, and
\vspace{-0.1cm}
\begin{eqnarray}
\Omega_{i,j}
& = &
\frac{ P_{i,j}}{N_j P_0} \tilde{d}_{i,j}^{-\alpha}
\label{eqn:omega}%
\end{eqnarray}
is the normalized power of $X_i$ at receiver $Y_j$ before despreading.

\vspace{-0.2cm}
\section{Outage Probability} \label{Section:Outage}
\label{Section:OutageProbability}
Let $\beta_j$ denote the minimum SINR required by $Y_j$ for reliable reception and $\boldsymbol{\Omega }_j=\{\Omega_{1,j},...,\Omega _{M,j}\}$ represent the set of normalized despread base-station powers received by $Y_j$.  An \emph{outage} occurs when the SINR falls below $\beta_j$.  As discussed subsequently, there is a relationship between the SINR threshold and the supported {\em rate} of the transmission.  Conditioning on $\boldsymbol{\Omega }_j$, the outage probability of mobile $Y_j$ is
\vspace{-0.1 cm}
\begin{eqnarray}
   \epsilon_j
   & = &
   P \left[ \gamma_j \leq \beta_j \big| \boldsymbol \Omega_j \right].
   \label{Equation:Outage1}
\end{eqnarray}
Because it is conditioned on $\boldsymbol{\Omega }_j$, the outage probability depends on the particular network realization, which has dynamics over timescales that are much slower than the fading.
By defining
\vspace{-0.3 cm}
\begin{eqnarray}
  \mathsf S = \sum_{k \in \mathcal G_j} \beta_j^{-1} g_{k,j} \Omega_{k,j}, \hspace{0.2 cm} Y_i =
  \frac{h}{G} g_{i,j} \Omega_{i,j}
\end{eqnarray}
\vspace{-0.7 cm}
\begin{eqnarray}
  \mathsf Z_j & = & \mathsf S  - \sum_{ i \notin \mathcal G_j }
  Y_i \label{eqn:z}
\end{eqnarray}
the conditional outage probability may be expressed as
\begin{eqnarray}
  \epsilon_j
  & = &
  P
  \left[
   \mathsf Z_j  \leq \Gamma^{-1} \big| \boldsymbol \Omega_j
  \right]
  = F_{\mathsf Z_j} \left( \Gamma^{-1} \big| \boldsymbol \Omega_j \right) \label{Equation:OutageCDF}
\end{eqnarray}
which is the cumulative distribution function (cdf) of $\mathsf Z_j$ conditioned on $\boldsymbol \Omega_j$ and evaluated at $\Gamma^{-1}$.
In the absence of interference, $F_{\mathsf S}\left(y\big| \boldsymbol \Omega_j\right)=F_{\mathsf Z_j}\left(y \big| \boldsymbol \Omega_j\right)$. Restricting the Nakagami parameters $m_{k,j}, \forall k \in \mathcal G_j$, to be integer-valued, but not necessarily identical, the conditional cdf of $\mathsf S$ is found in \cite{karagiannidis:2006} to be
\vspace{-0.3 cm}
\begin{eqnarray}
   F_{\mathsf S}\left(y\big| \boldsymbol \Omega_j\right)
  \hspace{-0.3 cm} &=& \hspace{-0.5 cm}
   \sum_{k \in \mathcal G_j} \hspace{-0.1 cm} \sum_{n=1}^{m_{k,j}}
   \left[1 - \hspace{-0.1 cm} \exp \left(\hspace{-0.1 cm} - \frac{\beta_j m_{k,j} y }{\Omega_{k,j}} \right) \hspace{-0.1 cm}  \sum_{\mu=0}^{n-1}\hspace{-0.1 cm}  \frac{\left(\beta_j m_{k,j} y \right)^\mu}{\Omega_{k,j}^\mu \mu!} \right] \nonumber \\
   & & \hspace{-0.7 cm}\Xi_{N_j} \hspace{-0.1 cm} \left( \hspace{-0.1 cm}k, n, \left\{ m_{q,j} \right\}_{\forall q \in \mathcal G_j},\left\{\frac{\Omega_{q,j}}{\beta_j m_{q,j}} \right\}_{\forall q \in \mathcal G_j } \right)
   \label{cdf_Naka_case2}
\end{eqnarray}
\begin{figure*}[ht]
\vspace{-0.5 cm}
\setcounter{equation}{14}
\begin{eqnarray}
 & &\hspace{-0.9 cm}\Xi_{L} \left( k, n, \left\{ r_q \right\}_{q=1}^{L},\left\{\eta_q \right\}_{q=1}^{L} \right) =
 \sum_{l_1=n}^{r_k} \sum_{l_2=n}^{l_1} \hspace{-0.1 cm} \cdots \hspace{-0.2 cm} \sum_{l_{L-2}=n}^{l_{L-3}}\hspace{-0.1 cm} \left[ \hspace{-0.1 cm} \frac{\left( -1 \right)^{R_L-r_k} \eta_k^n}{\prod_{h=1}^{L} \eta_h^{r_h}} \frac{\left( r_k+r_{1+u(1-k)}-l_1-1 \right)!}{\left( r_{1+u(1-k)} -1 \right)! \left( r_k - l_1\right)!}   \right. \nonumber \\
  & & \hspace{-0.9 cm} \left.
  \left( \frac{1}{\eta_j}-\frac{1}{\eta_{1+u(1-k)}} \right)^{l_1-r_k-r_{1+u(1-k)}} \frac{\left( l_{L-2}+r_{L-1+u(L-1-k)}-n-1 \right)!}{\left( r_{L-1+u(L-1-k)}-1\right)! \left(l_{L-2}-n\right)!}
  \left( \frac{1}{\eta_k} - \frac{1}{\eta_{L-1+u(L-1-k)}} \right)^{n-l_{L-2}-r_{L-1+u(L-1-k)}}  \right. \nonumber \\
  & & \hspace{-0.9 cm} \left.   \prod_{s=1}^{L-3} \frac{\left( l_s + r_{s+1+u(s+1-k)}-l_{s+1}-1\right)!}{\left( r_{s+1+u(s+1-k)}-1\right)! \left( l_s -l_{s+1}\right)!} \left(\frac{1}{\eta_k}-\frac{1}{\eta_{s+1+u(s+1-k)}}\right)^{l_{s+1}-l_s-r_{s+1+u(s+1-k)}}
\right]
   \label{Xi_L}.
\end{eqnarray}
\vspace{-0.10 cm}
{\hrulefill}
\end{figure*}
\setcounter{equation}{15}
The function $\Xi_{L} \left( k, n, \left\{ r_q \right\}_{q=1}^{L},\left\{\eta_q \right\}_{q=1}^{L} \right)$ is defined by (\ref{Xi_L}) at the top of the next page, where $u(x)$ is the step function and
\vspace{-0.3cm}
\begin{eqnarray}
R_L  &=&  \sum_{k=1}^{L} r_k.
\end{eqnarray}

A closed-form expression for $F_{\mathsf{Z}_j}(z | \boldsymbol \Omega_j)$ is given by (\ref{eqn_final_case1_Naka2}) at the top of the next page, with the proof is given in the Appendix.
\begin{figure*}[ht]
\vspace{-0.55 cm}
\setcounter{equation}{16}
\begin{eqnarray}
F_{\mathsf{Z}_j}(z \big| \boldsymbol \Omega_j)
 \hspace{-0.3 cm}& = &\hspace{-0.3 cm}
\sum_{k \in \mathcal G_j} \sum_{n=1}^{m_{k,j}} \Xi_{N_j} \hspace{-0.1 cm}\left( k, n, \left\{ m_{q,j} \right\}_{\forall q \in \mathcal G_j},\left\{\frac{\Omega_{q,j}}{\beta_j m_{q,j}} \right\}_{\forall q \in \mathcal G_j} \right) \left\{ 1 - \exp \left(\hspace{-0.1 cm}- \frac{\beta_j m_{k,j} z}{\Omega_{k,j}} \right) \sum_{\mu=0}^{n-1} \left(\frac{\beta_j m_{k,j} z }{\Omega_{k,j}} \right)^\mu \right. \vspace{-0.3 cm} \nonumber \\
\hspace{-0.3 cm} & & \hspace{-0.3 cm}
   \sum_{t=0}^\mu
\frac{z^{-t}}{\left( \mu - t \right)!} \hspace{-0.5 cm} \mathop{ \sum_{\ell_i \geq 0}}_{\sum_{i=0}^{M-N_j}\ell_i=t} \left.
\prod_{ i \notin \mathcal G_j }  \left[
 \ \frac{ \Gamma(\ell_i+m_{i,j}) }{ \ell_i! \Gamma( m_{i,j} ) }  \hspace{-0.1 cm}
\left( \frac{h \Omega_{i,j}}{G m_{i,j}} \right)^{\ell_i}
 \hspace{-0.1 cm} \left(
  \frac{\beta_j m_{k,j} }{\Omega_{k,j}} \frac{h \Omega_{i,j}}{G m_{i,j}} + 1
 \right)^{-(m_{i,j}+\ell_i)} \right] \right\}.
  \label{eqn_final_case1_Naka2}
\end{eqnarray}
\vspace{-0.1 cm}
\hrulefill
\vspace{-0.6 cm}
\end{figure*}
\setcounter{equation}{17}
Notice that (\ref{eqn_final_case1_Naka2}) reduces to (25) of \cite{torrieri:2012} when there is only one serving base station ($N_j=1$).

{\em Example $\#$ 1} In this example, as well in the rest of this paper, a \emph{distance-dependent
fading} model is assumed, where $m_{i,j}$ is set according to:
\vspace{-0.1 cm}
\begin{equation}
m_{i,j}=%
\begin{cases}
3 & \mbox{ if }\;||S_{j}-X_{i}||\leq r_{\mathsf{bs}}/2 \\
2 & \mbox{ if }\;r_{\mathsf{bs}}/2<||S_{j}-X_{i}||\leq r_{\mathsf{bs}} \\
1 & \mbox{ if }\;||S_{j}-X_{i}||>r_{\mathsf{bs}}%
\end{cases}%
.
\label{eqn:ddfading}
\end{equation}
The distance-dependent-fading model characterizes the situation where a mobile close to the base station is in the line-of-sight, while mobiles farther away are usually not.

\begin{figure}[t]
\centering
\includegraphics[width=8.75cm]{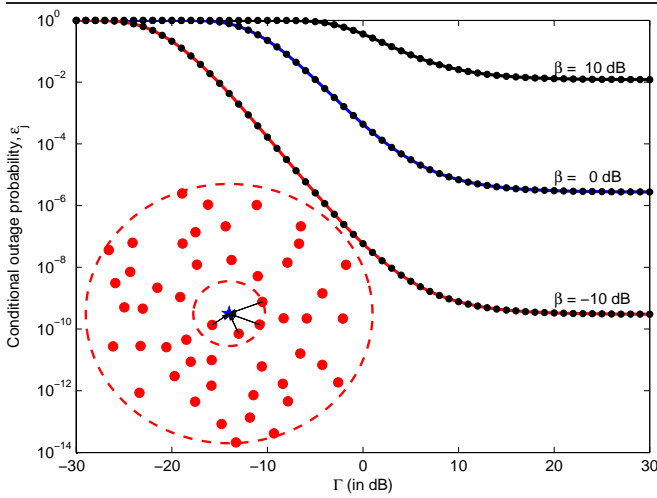}
\vspace{-0.75cm}
\caption{Conditional outage probability $\epsilon_{j}$ as a function of SNR $\Gamma$.  Analytical curves are solid while dots represent simulated values.  Top curve: $\beta = 10$ dB.  Middle curve: $\beta = 0$ dB.  Bottom curve: $\beta = -10$ dB. The network topology is shown in the inset.  The mobile is represented by the star at the center of the circular area, and the 50 base stations are shown as large filled circles. 
\label{Figure:Example1} }
\vspace{-0.7cm}
\end{figure}

The inset of Fig. \ref{Figure:Example1} shows an example network with $M=50$ base stations placed within a circular area and a single reference mobile located at the origin. The base-station locations are given by the large filled circles, and the mobile is represented by a star.  In the example, the mobile is served by the four closest base stations, which are the base stations located within a circle whose radius is one-fourth the radius of the network.  The four base stations serving the reference mobile are connected to it by arrows in the diagram. For this example, as well in the following, the fraction of power devoted to pilots is $f_p = 0.1$, the spreading factor is set to $G=16$ with chip factor $h=2/3$, and the propagation environment is characterized by a path-loss exponent $\alpha = 3$.

The outage probability was found by evaluating (\ref{eqn_final_case1_Naka2}) at $z=\Gamma^{-1}$ for three different values of $\beta_j$ and is shown in Fig. \ref{Figure:Example1}. Also shown are curves generated by simulation, which involved randomly generating the gamma-distributed  $\{g_{i,j}\}$.  The analytical and simulation results coincide, which is to be expected because  (\ref{eqn_final_case1_Naka2}) is exact.  Any discrepancy between the curves can be attributed to the finite number of Monte Carlo trials (one million trials were executed per SNR point).

\section{Policies}\label{Section:Policies}
A key consideration in the operation of the network is the
manner that base stations are associated with mobiles and the way that the total power $P_0$ transmitted by a base station is shared by the mobiles it serves.  These policies influence the rates provided to each user. This section discusses the cell-association, power-allocation, and rate-control policies.

\subsection{Cell Association}

Let $r_\mathsf{int}$ represent the radius of the cell {\em interior}.  Any mobile that is within an effective distance $r_\mathsf{int}$ of a base station will be served by just that base station, and mobiles beyond $r_\mathsf{int}$ will attempt to connect to a plurality of base stations out to some maximum distance $r_\mathsf{max}$, where $r_\mathsf{int} \leq r_\mathsf{max}$.  It follows that mobile $Y_j$ is served by those base stations $X_i$ whose indices are in the set
\vspace{-0.2 cm}
\begin{eqnarray}
\hspace{-0.1cm} \mathcal G_j
\hspace{-0.3cm} & = & \hspace{-0.3cm}
\begin{cases}
\underset{i}{\operatorname{argmin}}
\,
\left\{ \tilde{d}_{i,j}
\right\} & \hspace{-0.2cm} \mbox{if $ \underset{i}{\operatorname{\min}}
\, \left(\tilde{d}_{i,j}\right) \leq r_\mathsf{int}$} \\
\left\{ i : \tilde{d}_{i,j}< r_\mathsf{max}\right \}   & \hspace{-0.2cm} \mbox{if $ \underset{i}{\operatorname{\min}}
\, \left(\tilde{d}_{i,j}\right) > r_\mathsf{int}$}
\end{cases}
\label{eqn:connectivity}
\end{eqnarray}

Because there are only $G$ orthogonal spreading sequences per cell, the number of mobiles connected to $X_i$ is limited to $K_i \leq G$.  If there are $K_i > G$ mobiles in the set $\mathcal G_j$ defined by (\ref{eqn:connectivity}),  then some of these mobiles will either be refused service by $X_i$ or given service at a lower rate (through the use of an additional time multiplexing). In the following, we assume that the first $G$ mobiles whose effective distances are closest to $X_i$ are served, while the remaining $K_i - G$ are refused service and marked as having zero rate.

\subsection{Power allocation}
Base station $X_i$ allocates the power transmitted to mobile $Y_j$ according to the fractional power-control policy
\vspace{-0.1 cm}
\begin{eqnarray}
P_{i,j}=
P_0 \left( 1- f_p \right)\left[ \frac{1-\gamma}{K_i} + \gamma \tilde{d}_{i,j}^{\alpha} \left(\displaystyle{\sum_{j: Y_j \in \mathcal Y_i}} \tilde{d}_{i,j}^{\alpha}\right)^{-1} \right]
\label{Equation:Share}
\end{eqnarray}
where $0 \leq  \gamma \leq 1$ is the {\em fractional power-control factor}.  In this paper, the two extreme situations are considered: $\gamma = 0$, which corresponds to an {\em equal transmit power} (ETP) policy, and $\gamma = 1$, which corresponds to an {\em equal received power} (ERP) policy.

\subsection{Rate allocation}
Under the given power-allocation policy, the rate allocated to each user is selected to satisfy an outage probability.  In particular, the threshold $\beta_j$ of mobile $Y_j$ is selected such that the outage probability of mobile $Y_j$ satisfies the constraint $\epsilon_j \leq \hat{\epsilon}$. A constraint of $\hat{\epsilon} = 0.1$ is typical and appropriate for modern systems that use a hybrid automatic repeat request (HARQ) protocol.  For a given outage constraint, the SINR threshold depends on the modulation and coding scheme and receiver implementation.  For a given $\beta_j$, there is a corresponding transmission rate $R_j$ that can be supported, and typically only a discrete set of $R_j$ can be supported. Let $R_j = C(\beta_j)$ represent the relationship between $R_j$, expressed in units of bits per channel use (bpcu), and $\beta_j$. While the exact dependence of $R_j$ on $\beta_j$ can be determined empirically through tests or simulation,  we make the simplifying assumption when computing our numerical results that $C(\beta_j) = \log_2(1+\beta_j)$ corresponding to the Shannon capacity for complex discrete-time AWGN channels. This assumption is fairly accurate for systems that use capacity-approaching codes and a large number of code rates and modulation schemes, such as modern cellular systems, which use turbo codes with a large number of available modulation and coding schemes.  Once a $\beta_j$ is found that satisfies the outage constraint, the corresponding $R_j$ is found by using the function $R_j = C(\beta_j)$.


\section{Constrained Spatial Model}\label{Section:SpatialModel}
\vspace{-0.1 cm}
Under outage constraint $\hat{\epsilon}$, the performance of a given network realization is largely determined by the set of achieved rates $\{R_j\}$ of the $K$ users in the network.  We adopt a random spatial model, allowing the locations of the base stations and mobiles to vary, and determine the set of rates $\{R_j\}$ for each random realization. Unlike other work, which models the spatial locations of the base stations according to a PPP, we adopt a {\em constrained} spatial model, which requires a minimum spacing among base stations.  In the model, the $M$ base stations and $K$ mobiles are placed on a disk of radius $r_\mathsf{net}$, so that the network area $A_\mathsf{net} = \pi r^2_\mathsf{net}$.  An {\em exclusion zone} of radius $r_{bs}$ surrounds each base station, and no other base stations are allowed within this zone.  Similarly, an exclusion zone of radius $r_{m}$ surrounds each mobile, and no other mobiles are allowed within a placed mobile's exclusion zone. Furthermore, each mobile is at least distance $r_{m}$ away from any base station.

Let the random variable $R$ represent the rate of an arbitrary user in a realization of the spatial model.  The statistics of $R$ can be found for a given set of $\{M,K,r_\mathsf{net},r_\mathsf{bs},r_\mathsf{m},\sigma_s\}$ using a Monte Carlo approach as follows.  Draw a realization of the network by placing $M$ base stations and $K$ mobiles within the disk of radius $r_\mathsf{net}$ according to a uniform clustering model.
  Compute the path loss from each base station to each mobile, applying randomly generated shadowing factors if shadowing is present.  Determine the set of mobiles associated with each base station, applying the cell association policy.  At each base station, apply the power-allocation policy to determine the power it transmits to each mobile that it serves.  By setting the outage equal to the outage constraint, invert (\ref{eqn_final_case1_Naka2}) to determine the SINR threshold for each mobile in the cell.  By applying the function $R_j=C(\beta_j)$, find the rate of the mobile.  Repeat this process for a large number of network realizations, all with the same values of $\{M,K,r_\mathsf{net},r_\mathsf{bs},r_\mathsf{m},\sigma_s\}$.

Let $\mathbb E[R]$ represent the mean value of the variable $R$, which can be found by numerically averaging the values of $R$ obtained using the procedure described in the previous paragraph.  While $E[R]$ is a useful metric, it does not account for the loss in throughput due to the inability to successfully decode during an outage, and it does not account for the spatial density of transmissions.   These considerations are taken into account by the \emph{area spectral efficiency}, which is defined as \cite{weber:2010}
\vspace{-0.3cm}
\begin{eqnarray}
  \mathcal A
  & = &
  \lambda
  \left( 1 - \hat{\epsilon} \right)
  \mathbb E[R]
  \label{eqn:tc}
\end{eqnarray}
where $\lambda = K/A_\mathsf{net}$ is the density of transmissions in the network, or equivalently, the number of active receivers per unit area.  
\vspace{-0.2cm}
\section{Numerical Results}\label{Section:Results}
\vspace{-0.1 cm}
As an example, consider a network with $M=50$ base stations placed in a network of radius $r_\mathsf{net} = 2$ with base-station exclusion zones of radius $r_\mathsf{bs} = 0.25$.  A variable number $K$ of mobiles are placed within the network using exclusion zones of radius $r_\mathsf{m} = 0.01$.  The outage constraint is set to $\hat{\epsilon} = 0.1$. The SNR is set to $\Gamma = 10$ dB.


\begin{figure}[t]
\centering
\includegraphics[width=8.75cm]{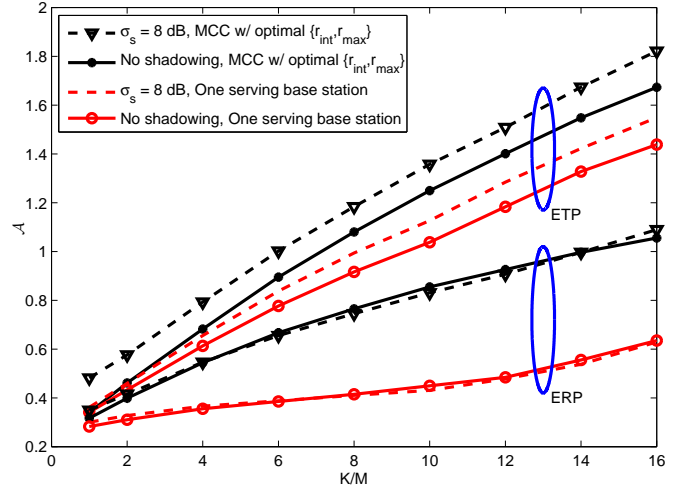}
\vspace{-0.75cm}
\caption{ Area spectral efficiency  as a function of $K/M$ with ERP and ETP policy, for both a shadowed ($\sigma_s = 8$ dB) and unshadowed environment.
\label{Figure:TC_KM} }
\vspace{-0.35cm}
\end{figure}

\begin{figure}[t]
\centering
\includegraphics[width=8.75cm]{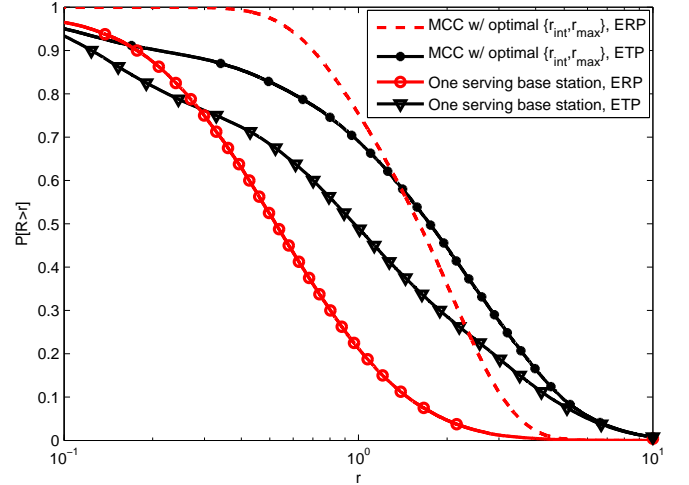}
\vspace{-0.75cm}
\caption{ Ccdf of $R$ for ERP and ETP policy for a half loaded system $(K/M=8)$ in shadowing ($\sigma_s = 8$ dB). \label{Figure:CcdfRate}  }
 \vspace{-0.65cm}
\end{figure}


Fig. \ref{Figure:TC_KM} shows the area spectral efficiency as a function of the ratio $K/M$, of both ERP and ETP policies in an unshadowed
environment as well as in the presence of shadowing. Results are shown for an MCC system using the value of  $\{r_\mathsf{int},r_\mathsf{max}\}$ that maximizes area spectral efficiency, as well as a conventional system that limits each mobile to being served by no more than one base station.
The figure shows the potential advantage of using MCC and the superior area spectral efficiency of the ETP policy. The figure shows that shadowing actually improves the performance, especially when ETP is used. This is because shadowing can sometimes cause the effective distance between a mobile and its serving base station(s) to decrease, while its effective distance to the interfering base stations is increased.  Even though this scenario does not occur very often, when it does the rate can be significantly increased, causing the average to increase.


The fairness of a particular power allocation policy can be further visualized by plotting the complementary cdf of $R$, which is the probability that $R$ exceeds a threshold $r$; i.e., $P[R>r]$.  Fig. \ref{Figure:CcdfRate} shows the ccdf of $R$ in presence of shadowing ($\sigma_s = 8$ dB), using the ETP and ERP policies when the system is half loaded $(K/M=8)$.  The cumulative cdf curves for the ERP policy are steeper than they are for the ETP policy, indicating less variability in the provided rates. The lower variability in rate corresponds to improved fairness.


\vspace{-0.1 cm}
\section{Conclusion} \label{Section:Conclusion}
This paper has presented a new approach for modeling and analyzing the performance of multi-cell downlink cooperation. While the analysis and the model have been used to study a DS-CDMA cellular downlink, this work could be extended to analyze different types of access, such as orthogonal frequency-division multiple access (OFDMA).  The analysis combines a new outage probability expression, which is exact for a given network realization, with a constrained random spatial model, which allows the statistics to be determined for a class of network topologies.   A fractional power-control policy and outage-constrained rate-control policy are considered, and numerical results are provided for the extremes of ETP and ERP.  It is found that ETP is more favorable in terms of the area spectral efficiency, while ERP is more fair.
\vspace{-0.1 cm}
\appendix
The cdf of $\mathsf Z$ can be computed as follows:
\vspace{-0.2 cm}
\begin{eqnarray}
   F_{\mathsf{Z}_j}(z \big| \boldsymbol \Omega_j)
   \hspace{-0.3 cm}& = &\hspace{-0.3 cm}
   P [ \mathsf Z_j < z ]
   =
   P \left[ \mathsf S < z + \sum_{ i \notin \mathcal G_j } \mathsf Y_i \right] \nonumber \\
   \vspace{-0.1 cm}
   \hspace{-0.3 cm}& = &\hspace{-0.3 cm}
   \underset{\mathbb R^{M-N_j}}{ \int ... \int }
f_{\mathbf Y}( \mathbf y) F_{\mathsf S} \left( z + \sum {y_i} \right) d \mathbf y \label{eqn1}
\vspace{-0.2 cm}
\end{eqnarray}
where the outer integral is over $\left(M-N_j\right)$-dimensional real space and $f_{\mathbf Y}( \mathbf y)$ is the joint pdf of the vector $\mathbf Y$, where each element is gamma-distributed with Nakagami parameter $m_{i,j}$ with pdf
\vspace{-0.2 cm}
\begin{eqnarray}
   f_{\mathsf{Y}_i}(y) = \left( \frac{G m_{i,j}}{h \Omega_{i,j}} \hspace{-0.1 cm} \right)^{m_{i,j}}  \frac{y^{m_{i,j}-1 }}{\Gamma\left(m_{i,j}\right)}\hspace{-0.1 cm} \exp\left( \hspace{-0.1 cm}- \frac{ G m_{i,j}y}{h \Omega_{i,j}}\right)\hspace{-0.1 cm} u(y).
    \label{pdf_yi}
\end{eqnarray}
Substituting (\ref{cdf_Naka_case2}) into (\ref{eqn1}) and performing a few manipulations yields
\vspace{-0.4 cm}
\begin{eqnarray}
F_{\mathsf{Z}_j}(z \big| \boldsymbol \Omega_j)
  \hspace{-0.3 cm} &=& \hspace{-0.4 cm}
   \sum_{k \in \mathcal G_j} \hspace{-0.1 cm} \sum_{n=1}^{m_{k,j}}
   \hspace{-0.1 cm} \Xi_{N_j} \hspace{-0.1 cm} \left( \hspace{-0.1 cm}k, n, \left\{ m_{q,j} \right\}_{\forall q \in \mathcal G_j}\hspace{-0.1 cm},\hspace{-0.1 cm}\left\{\hspace{-0.1 cm}\frac{\Omega_{q,j}}{\beta_j m_{q,j}}\hspace{-0.1 cm} \right\}_{\forall q \in \mathcal G_j} \right)
 \nonumber \\
 \hspace{-0.3 cm}& &\hspace{-0.8 cm} \left\{
   1\hspace{-0.1 cm} - \hspace{-0.1 cm} \exp \left(\hspace{-0.1 cm} - \frac{\beta_j m_{k,j} z }{\Omega_{k,j}} \right) \hspace{-0.1 cm}  \sum_{\mu=0}^{n-1}\hspace{-0.1 cm}  \frac{\left(\beta_j m_{k,j} z \right)^\mu}{\Omega_{k,j}^\mu \mu!}\hspace{-0.1 cm}\underset{\mathbb R^{M-N_j}}{ \int\hspace{-0.1 cm} ... \hspace{-0.1 cm}\int } f_{\mathbf Y}( \mathbf y)
   \right. \nonumber \\
 \hspace{-0.3 cm}& &\hspace{-0.8 cm}  \left.
 \exp \left(-\frac{\beta_j m_{k,j}}{\Omega_{k,j}}  \sum y_i \right) \hspace{-0.2 cm} {\left( 1 + z^{-1} \sum_{ i \notin \mathcal G_j } y_i \right)}^\mu
\hspace{-0.2 cm}  d \mathbf y \right\}\hspace{-0.1 cm}.
\label{eqn3_Naka2}
\end{eqnarray}
Since $\mu$ is an integer, the binomial theorem gives the series expansion:
\vspace{-0.3 cm}
\begin{eqnarray}
{\left( 1 + z^{-1} \sum_{ i \notin \mathcal G_j } y_i \right)}^\mu
& = &
\sum_{t=0}^\mu
\binom{ \mu }{ t } z^{-t} \left( \sum_{ i \notin \mathcal G_j } y_i \right)^t \hspace{-0.2cm}.
\label{eqn4_Naka2}
\end{eqnarray}
A multinomial expansion yields
\vspace{-0.2 cm}
\begin{eqnarray}
\left( \sum_{ i \notin \mathcal G_j } y_i \right)^t
& = &
t! \mathop{ \sum_{\ell_i \geq 0}}_{\sum_{i=0}^{M-N_j}\ell_i=t}
\left( \prod_{ i \notin \mathcal G_j } \frac{ y_i^{\ell_i} }{\ell_i !} \right).\label{eqn5_Naka2}
\end{eqnarray}
Substituting (\ref{eqn4_Naka2}) and (\ref{eqn5_Naka2}) into (\ref{eqn3_Naka2}), performing a few manipulations, and assuming that the interfering channels fade independently, we obtain
\begin{eqnarray}
F_{\mathsf{Z}_j}(z \big| \boldsymbol \Omega_j)
  \hspace{-0.3 cm} &=& \hspace{-0.4 cm}
   \sum_{k \in \mathcal G_j} \hspace{-0.1 cm} \sum_{n=1}^{m_{k,j}}
   \hspace{-0.1 cm} \Xi_{N_j} \hspace{-0.1 cm} \left( \hspace{-0.1 cm}k, n, \left\{ m_{q,j} \right\}_{\forall q \in \mathcal G_j}\hspace{-0.1 cm},\hspace{-0.1 cm}\left\{\hspace{-0.1 cm}\frac{\Omega_{q,j}}{\beta_j m_{q,j}}\hspace{-0.1 cm} \right\}_{\forall q \in \mathcal G_j} \right)
 \nonumber \\
 \hspace{-0.3 cm}& &\hspace{-0.8 cm} \left\{
   1\hspace{-0.1 cm} - \hspace{-0.1 cm} \exp \left(\hspace{-0.1 cm} - \frac{\beta_j m_{k,j} z }{\Omega_{k,j}} \right) \hspace{-0.1 cm}  \sum_{\mu=0}^{n-1}\hspace{-0.1 cm}  \frac{\left(\beta_j m_{k,j} z \right)^\mu}{\Omega_{k,j}^\mu \mu!}\hspace{-0.1 cm}\sum_{t=0}^\mu
\frac{z^{-t}}{\left( \mu - t \right)!}  \right. \nonumber \\
\vspace{-0.2 cm}
 \hspace{-0.3 cm}& &\hspace{-1.3 cm}
\mathop{ \sum_{\ell_i \geq 0}}_{\sum_{i=0}^{M-N_j}\ell_i=t}
\hspace{-0.2 cm} \left.\prod_{ i \notin \mathcal G_j }  \int_{\mathbb R} \hspace{-0.1 cm}
 \exp \left( \hspace{-0.1 cm} - \frac{\beta_j m_{k,j} y}{\Omega_{k,j}}\right) \hspace{-0.1 cm} \frac{ y^{\ell_i} }{\ell_i !}
f_{ Y_i}( y) d y \right\} \hspace{-0.1 cm}.
\label{eqn8_Naka2}
\end{eqnarray}
Substituting (\ref{pdf_yi}), the integral in (\ref{eqn8_Naka2}) is
\begin{eqnarray}
& & \hspace{-1.0 cm} \int_{\mathbb R} \exp \left( - \frac{\beta_j m_{k,j} y}{\Omega_{k,j}}\right)  \frac{ y^{\ell_i} }{\ell_i !}  f_{\mathsf Y_i}(y) d y
  = \nonumber \\
& & \hspace{-1.0 cm}  \frac{ \Gamma(\ell_i+m_{i,j}) }{ \ell_i! \Gamma( m_{i,j} ) }
 \left( \frac{h \Omega_{i,j}}{G m_{i,j}} \right)^{\ell_i}
  \left(
  \frac{\beta_j m_{k,j} }{\Omega_{k,j}} \frac{h \Omega_{i,j}}{G m_{i,j}} + 1
 \right)^{-(m_{i,j}+\ell_i)}
 \label{eqn9_Naka2}.
\end{eqnarray}
Finally, by substituting (\ref{eqn9_Naka2}) into (\ref{eqn8_Naka2}), (\ref{eqn_final_case1_Naka2}) is obtained.

\bibliographystyle{ieeetr}
\bibliography{gb2013}

\balance
\end{document}